**Long Title** Comparison of Aerosol Emissions during Specific Speech Tasks

**Short Title** Aerosol Emissions during Speech Tasks

Miriam van Mersbergen[1,3,4], Jeffrey Marchetta[2], Daniel Foti[2], Eric Pillow[2], Apartim Dasgupta[2], Chandler Cain[2], Stephen Morvant[3]

[1]School of Communication Sciences and Disorders, University of Memphis, Memphis, Tennessee, United States of America

[2]Department of Mechanical Engineering, University of Memphis, Memphis, Tennessee, United States of America

[3]Department of Otolaryngology, Head and Neck Surgery, University of Tennessee Health Sciences Center, Memphis, Tennessee, United States of America

[4]Institute for Intelligent Systems, University of Memphis, Tennessee, United States of America

Corresponding author:

Miriam van Mersbergen, Ph.D., CCC-SLP
Assistant Professor
School of Communication Sciences and Disorders
The University of Memphis
4055 North Park Loop
Memphis, Tennessee 38152
Miriam.van.Mersbergen@memphis.edu

**Abstract**

The study of aerosols and droplets emitted from the oral cavity has become increasingly important throughout the COVID-19 pandemic. Studies show particulates emitted while speaking were generally much smaller compared to coughing or sneezing. However, recent investigations revealed that they are large enough to carry respiratory contagions. Although studies have shown that particulate emissions do indeed occur during speech, to date, there is little information about the relative contribution of different speech sounds in producing particle emissions. This study compares airborne aerosol generation in participants producing isolated speech sounds: fricative consonants, plosive consonants, and vowel sounds. While participants produced isolated speech tasks, a planar beam of laser light, a high-speed camera, and image software calculated the number of particulates detected overtime. This study compares airborne aerosols emitted by human participants at a distance of 2.54 cm between the laser sheet and the mouth and reveals statistically significant increases in particulate counts over ambient dust distribution for all speech sounds. Vowel sounds were statistically greater than consonants, suggesting that mouth opening, as opposed to place of vocal tract constriction or manner of sound production, might be the primary influence in the degree to which particulates become aerosolized during speech. Results of this research will inform boundary conditions for computation models of aerosolized particulates during speech.

**Introduction**

The study aims to quantify aerosols and droplets that range from one to one hundred micrometres in diameter emitted during isolated speech tasks. Recent studies have found that two of the three primary routes of human-to-human transmission of COVID-19 include the droplet and airborne routes.[1–4] The droplet route involves those larger particles that are ejected at a close enough range to contact another person's mouth or nose directly, while the airborne route involves the inhalation of aerosolized droplet nuclei which can become entrained in the airflow after their initial expulsion.[5] Regardless of the size of the particle, droplet or airborne, it is difficult to determine if those particles carry a viral load. As such, merely measuring particle count does not unequivocally determine if those particles carry viral loads. Despite this, measuring particulate count is a reasonable method to estimate the *potential* for viral transmission.

Aerosols, usually defined as particles less than 10 micrometres in diameter, are a subject of particular interest because virus-carrying aerosols can stay airborne and travel through a space for hours, posing an extended risk of infection.[1] A study published in 2009 demonstrated that exhaled particles carry multiple pathogens, including influenza.[6] However, if a proportion of exhaled particles that are likely to carry a viral load is known, one still must be able to determine how many particles a person encounters before calculating the risk of transmission for a given scenario.[7] Thus, based on reasonable estimates of viral load in a given particle, a meaningful investigation of the risk of viral exposure for an activity should consider the number and size of particles a person encounters.[2,8–10]

Some of the best and most well-known investigations in this topic are focused on investigating the airflow that originates at the oral and nasal cavity.[11] Techniques used to observe the airflow usually include shadow-graphing or Schlieren imaging to collect optical data.[12] This kind of imaging is non-intrusive to the expiration event, allowing expulsion of air flow into a free field. Thus, the data are ecologically relevant for everyday scenarios. This imaging technique lends itself to studies that focus on abrupt events like coughing and sneezing in which documented airspeed during a cough may be of importance.[13] Other studies that apply this technique to speaking and regular breathing show airflow events during speech. Unfortunately, these methods do not provide data about the particles expelled by the airflows they investigate, which leaves estimations of viral load yet to be calculated.

Other studies focus exclusively on the particulate matter that is expelled during a cough, a sneeze, or regular breathing.[11,14,15] The number, size, and potential pathogen load have been investigated and documented in various studies.[6] However, the techniques used for these studies are usually intrusive and involve collection of the particles in a funnel,[16–18] tube,[19] or box[20] as soon as they leave the participant's oral or nasal cavity, interrupting the airflows that occur naturally in free space. Additionally, although these studies quantify the number of potential virus laden particles, visualization of the behaviour of particles as they disperse in the air remains under investigation.[21]

Formulating effective protection from viral exposure requires information on how individuals contribute to the sum of virus laden particles in the environment, and how to protect themselves from subsequent exposure to increased viral load. As such, understanding the nature of how particulates and the ensuing aerosolized infectious load are emitted from the oral cavity and interact with the environment requires in-depth study. Previous work that investigated particle emission in speech has used connected speech sounds in sentences,[16,19–21] singing phrases,[19,21,22] oral breathing,[15] coughing, and sneezing,[11] and has shown that this line of research requires controlled investigations of multiple factors contributing to particulate emissions and its resultant behavior. However, such studies miss key details such as the changes in speaking jet angles with different sounds.[23] Recently, a meta-study[24] found experiments on aerosolization with speech sounds used relatively few numbers of human participants (typically ~1).[20,23,25–32] The meta-study focuses mainly on aerosol production with plosives and identifies high particle production for plosives.

Speech and singing are inherently complicated activities, involving the interaction of aerodynamic forces with the mucosal lining of the vocal folds and vocal tract.[33] Vortices induced by vocal fold vibration and vocal tract constriction[34] have the potential to shear mucus-lined walls in the oral cavity and stretch saliva, which then distribute these mucous droplets into the air as they are expelled from the mouth. To better understand the conditions that contribute to the expulsion of particulate matter, investigating the effect of vocal tract constriction at various places along the path from the vocal folds to the exit from the oral cavity appears to be a good starting point. In essence, investigations seeking to understand the relative contributions individuals have in producing particulates should include parsing out the specific speech gestures that produce increased particulates, identifying conditions where particulates might stay aerosolized, while controlling for physiological differences in participants' physical constitution.

This larger pilot study aims to give a visual representation of the particles emitted from the nose and mouth, controlling for speech sound production and physical condition, while also providing quantitative conclusions about particle count, size, distribution, and dispersion at three different close-range distances from the source. A careful process of image collection and algorithmic analysis for data collection allows both objectives to be accomplished. The unique experimental setup of this study allows participants to be safely positioned near a light source powerful enough to illuminate the particles in a free field and capture images of them with a high-speed camera for analysis. The images and image sequences can provide a better understanding of the behavior of the particles, while the numerical and statistical data can be used to verify computational models designed to predict the motion of the particles under specific speech conditions.

The purpose of this current study was twofold: first, to test the feasibility of measuring particulate emissions from humans in a free field and second, to determine if individual speech sounds produced particulate emissions at varying levels. Results from both research questions will be factored into a larger scale experimental study with the purpose of developing computational models of aerosol and droplet behaviour.

**Materials and Methods**

This investigation was approved by the Institutional Review Board at the University of Memphis and the University of Tennessee Health Science Center. Participants included 9 females (average age 30.2 years, range 24-47 years), and 11 males (average age 39.6 years, range 28-63 years) recruited through social media and word-of-mouth. The study included 4 female and 4 male singers. All participants consented to participate. The final analysis omitted three participants, one female and two males due to technical malfunction.

   A. Experiment Setup

To meet the aims of the research study, an experiment was designed with the goal of quantifying and comparing particle counts at a close-range distance in a free field while participants performed isolated speech sounds. Figure 1 illustrates the setup of the proposed experiment which included the laser, high-speed camera, and participant head enclosure mounted on an optical table. Each participant performed isolated speech sounds while seated with their head placed in a small enclosure that consisted of a frame with a solid front panel with an opening for the nose and mouth. The laser and camera were positioned on a table, and the entire setup was enclosed in a 3m x 3m x 3.7m sealed canopy tent as shown in figure 2. A dehumidifier was used to maintain a relative humidity of 35% in the enclosure. An 11 $m^3$/min four-layer HEPA filtration unit was used to reduce the ambient dust in the space before each data collection scenario. The filter was turned off during testing to mitigate air movement so as to not interfere with the behavior of the aerosol expelled by the participant. Laser sheet imaging was used for data collection. Big Sky Ultra, dual pulsed, neodymium yttrium-aluminium garnet (ND-YAG), 532 nm lasers were used for the study. Each laser was pulsed at rate of 30 Hz and operated at their maximum power of 30mJ. The beam passed through a beam-spreader lens to transform the shape of the laser into a vertical sheet which was approximately 3mm thick. This laser sheet passed in front of an opening in a vertical board, which the participant sat behind. As the participant performed the vocal exercises, the aerosols scattered the laser light as they passed through the illuminated plane, and the light was scattered into the lens of the camera. A 1280 x 1024 pixel PCO 1200 HS camera captured grey scale images of all the aerosols that passed through the plane of the laser at 40 frames per second.

**Figure 1. Model of camera, laser, and enclosure frame setup**

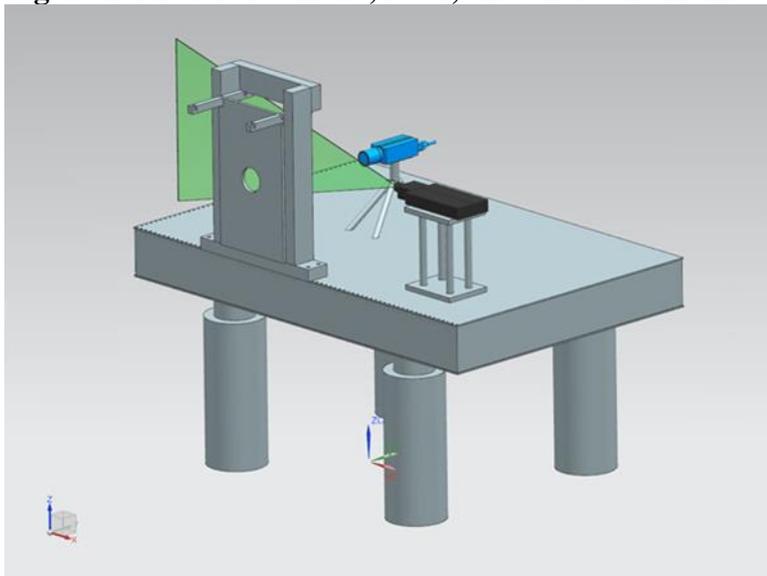

The participant head frame and cloth enclosure shown in figure 2 and figure 3 were designed to increase participant comfort while reducing reflective light from the laser sheet. The front of the enclosure was constructed of a solid wood frame, and its main purpose was to ensure the source of particulates, namely, the participant's oral and nasal cavity, stayed at a consistent distance from the laser. The frame was adjusted such that the participant was seated a consistent distance away from the fixed laser sheet at 2.5cm. All proper personal protective equipment, including laser safety goggles, were provided for the participant and the technical operator in the lab.

**Figure 2. Canopy tent and experimental environment.**

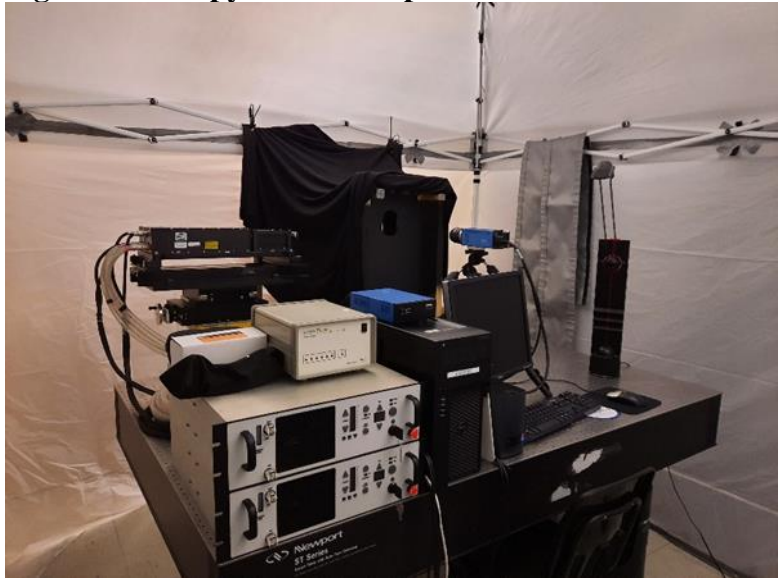

**Figure 3. Participant position behind the laser barrier.**

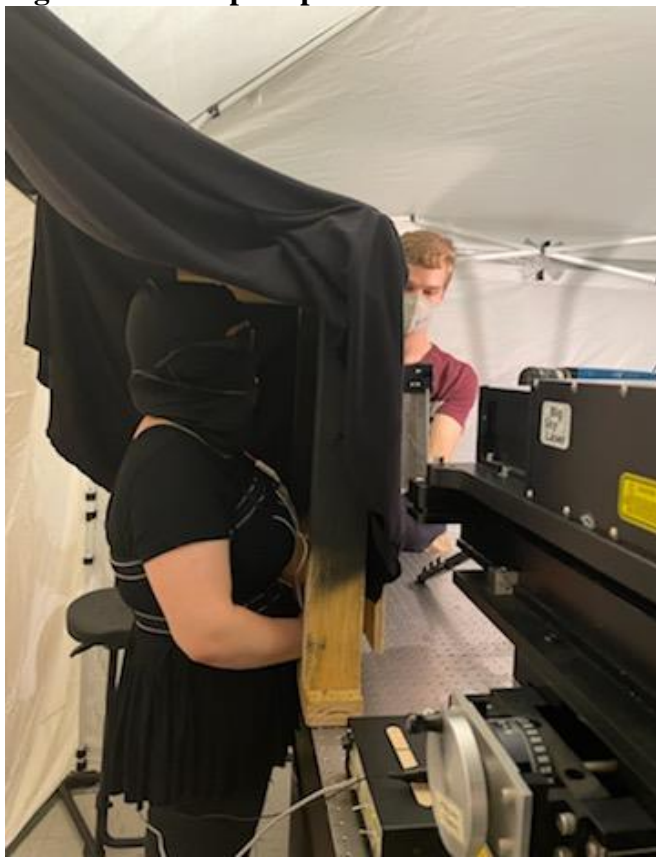

B. Experiment Procedure

Prior to particulate data collection, physiological measurements including body mass index, spirometry, subglottal pressure, glottal air flow rates, and glottal contact area were collected to serve as control variables. This de-identified data will not be presented here but may be used as pilot data in future investigations on individual factors of physical constitution that affect levels of aerosol expiration employing a larger number of participants.

Before a participant entered the laser laboratory, a speech language pathologist trained participants in the speech tasks that were elicited during data collection. Elicited tasks included isolated productions of plosive and fricative consonants (/p, b, k, g, ʃ, ʒ/), vowels (/a, i, u/), singing, and oral breathing. Speech and singing tasks were completed at two different loudness levels[35] and voice speech tasks included low pitch, based on average speaking fundamental frequency, and high pitch, based on the 80% mark of the total range. Speech tasks were recorded using a Zoom H6 Handy recorder (Tokyo, Japan) at 48 kHz sampling rate and 24-bit depth with an AKG C520 headworn cardioid condenser microphone (Vienna, Austria) at a $45^0$ angle 10 cm from the mouth center. Each production fell between 53 dB SPL and 72 dB SPL. For this initial feasibility analysis, speech tasks and phonetic classifications being compared in this study are presented in Table 1. For this analysis, all tasks were at normal pitch and normal loudness and counterbalanced across participants for place and manner. Cognate pairs were recorded within the same trial beginning with voiceless phonemes.

| IPA symbol | Written/Spoken | Place of Articulation | Manner | Voicing | |
|---|---|---|---|---|---|
| p | "p" as in pat | Front | Plosive | Unvoiced | Cognate Pair |
| b | "b" as in bat | Front | Plosive | Voiced | |
| k | "c" as in kale | Back | Plosive | Unvoiced | Cognate Pair |
| g | "g" as in gale | Back | Plosive | Voiced | |
| ʃ | "sh" as in motion | Mid-palatal | Fricative | Unvoiced | Cognate Pair |
| ʒ | "zh" as in vision | Mid-palatal | Fricative | Voiced | |
| i | "ee" as in heat | Front | Sonorant | Voiced | |
| a | "ah" as in hot | Middle | Sonorant | Voiced | |
| u | "oo" as in hoot | Back | Sonorant | Voiced | |

**Table 1. List of phonemes tested according to place, manner, and voicing characteristics.**

As a control, prior to the participant entering the enclosure for data collection, the laser was pulsed, and the camera captured images of the ambient dust in the space. Once in the laboratory, participants were positioned on an adjustable seat to line up their mouth and nose with the opening in the enclosure as previously described. As they performed the speech tasks, the laser illuminated the aerosols, and the camera captured a time sequence of images. An audio recording and a sound level reading were also collected and synchronized to the high-speed camera to determine the specific time frames for analysis. The camera recorded for approximately 8 seconds while the participant completed the sequence of speech tasks. At 40 fps, each video sequence produced approximately 300-400 images.

Following data collection, the participants were given questionnaires about demographics. Once these questionnaires were completed, they were thanked and paid for their participation. The entire study lasted about 2.5 hours.

B. Data Reduction and Analysis

An image post-processor, Image J, analysed the sequence images for particle counts and sizes captured for the control and the speech tasks. Audio waves and particle count were visually and temporally synchronized using an external clapperboard. Only particle counts that occurred within the time frame of the audio signal were included in the analysis. The synchronization of the audio file with the particle counts increased the confidence that a statistically significant increase in particle count was likely a consequence of the speech task. Examples from a time sequence of raw images for a participant vocalizing the three instances of the front-plosive consonant /p/ at a distance of 2.5cm are shown with the audio synchronization in figure 4.

**Figure 4. Audio synchronization for particle counts /p/.**

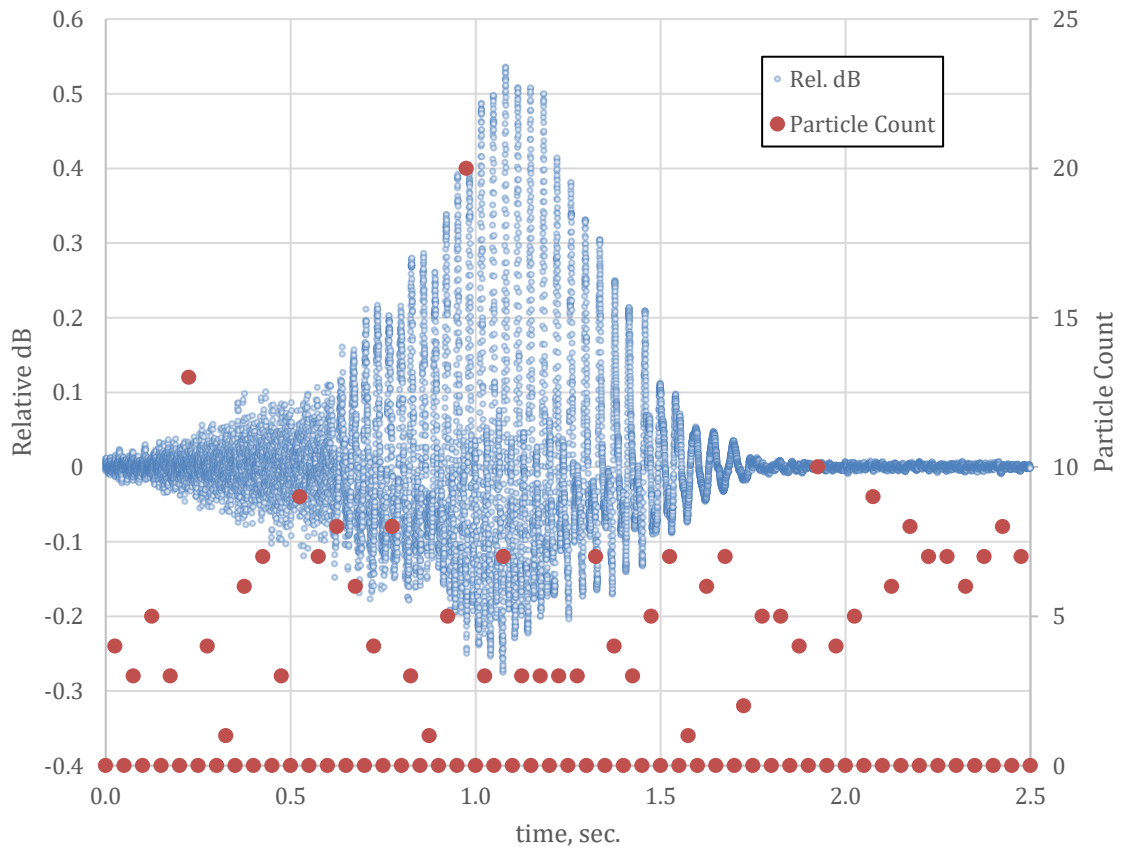

In answer to the first question about the feasibility of measuring particulate emissions from humans in a free field that could then be used in a computational model, the synchronized audio recordings were used to confirm the particle counts associated with specific speech tasks. For each confirmed phoneme, the maximum particle count was determined and aggregated into one data set. For each phoneme, the aggregated particle count was compared against the average background dust sets, which was aggregated separately. To analyse the second question, which was to determine if individual speech sounds produced particulate emissions at varying levels, a separate repeated-measures analysis of variance (ANOVA; SPSS v.25, IBM) was used to compare voicing (unvoiced vs voiced) place (front, mid, and back assessed separately for consonants and vowels), manner (fricatives vs plosives), and sequence (first, second, third trial). Alpha levels were set at *p*=.05 for each ANOVA. Linear and quadratic analysis were then applied to each ANOVA as to assess the relationship between the individual factors. No post hoc analysis on specific conditions were completed.

**Results**

A. Signal to Noise Assessment

Nine speech sounds were analysed using a one-way ANOVA against the background dust: two front-plosives (/p, b/), two back-plosives (/k. g/), two mid-palatal fricatives (/ʃ, ʒ/), and three vowels (/a, i, u/). All utterances were statistically significant at a confidence of 95% compared to the background dust. The count, mean, variance, confidence, and p-values are presented in Table 2. As shown in Figure 5, the background dust did not vary significantly across participant runs.

| Phoneme | Mean | Std. Deviation | 95% Confidence | ANOVA of Dust vs Phoneme |
|---|---|---|---|---|
| *p* | 6.76 | 4.116 | 1.13 | F(2,51) = 357.43, *p* < .001 |
| *b* | 6.12 | 2.389 | 0.66 | F(2,51) = 286.44, *p* < .001 |
| *k* | 6.53 | 3.466 | 0.95 | F(2,51) = 331.53, *p* < .001 |
| *g* | 6.43 | 3.727 | 1.02 | F(2,51) = 318.13, *p* < .001 |
| *ʃ* | 7.94 | 7.806 | 2.14 | F(2,51) = 477.47, *p* < .001 |
| *ʒ* | 7.06 | 4.662 | 1.28 | F(2,51) = 391.96, *p* < .001 |
| *a* | 7.12 | 2.118 | 1.01 | F(2,17) = 139.51, *p* < .001 |
| *i* | 6.29 | 1.649 | 0.78 | F(2,17) = 103.58, *p* < .001 |
| *u* | 6.65 | 2.090 | 0.99 | F(2,17) = 118.26, *p* < .001 |
| *Dust* | 5071 | 1.1889 | 0.06 | |

**Table 2. Descriptive and ANOVA statistics for particle counts compared to dust**

**Figure 5. Graphical display of particles generated from phoneme production compared to background dust.**

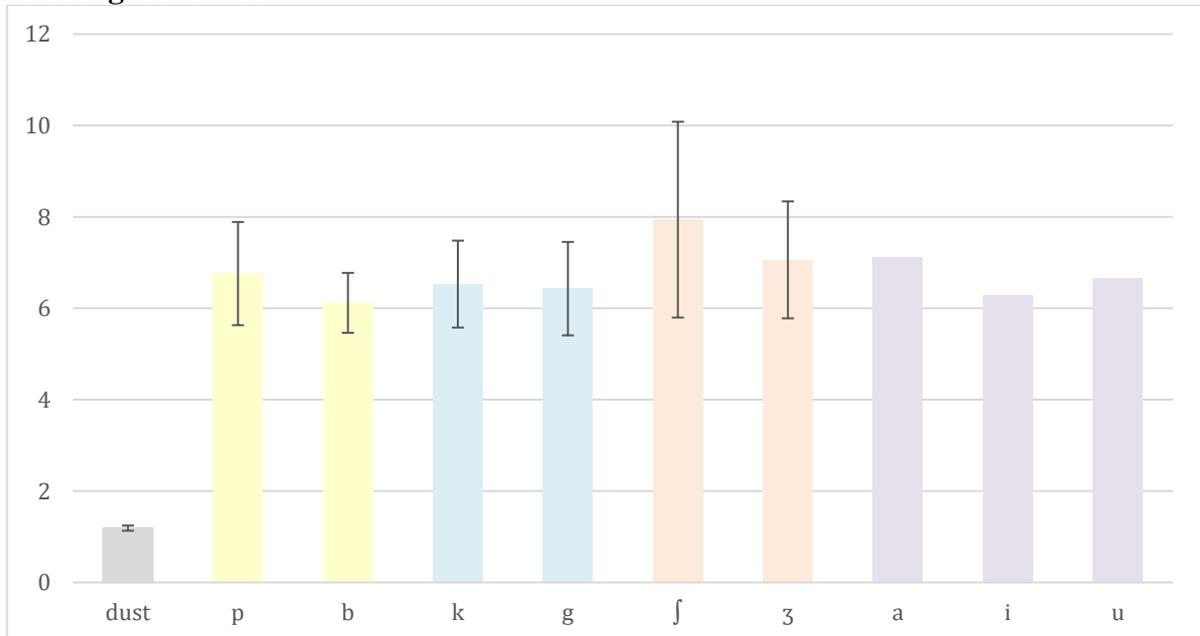

Error bars represent 95% confidence interval.

B. Speech Sound Assessment

In answer to the question if particulate count differed between specific linguistic categories, findings were mixed. Between voiced and unvoiced consonants, a repeated-measures ANOVA showed no significant main effect, $F(1, 16) = .938$, $p = 0.347$. Similarly, particulate count did not differ according to consonant placement, $F(2, 32) = .484$, $p = 0.540$, or between plosive and fricative consonants, $F(1, 16) = .554$, $p = 0.467$.

However, particulate count differed according to sequence of utterance, F(2, 32) =8.747, p = 0.050. Additionally, the linear contrast for the sequence of utterance reached significance, F(1,16) =6.007, p = 0.026, suggesting that particulates increased in number as utterances increased in number. Finally, particulate count differed according to vowel placement, F(2, 32) = 3.637, p = 0.038. Furthermore, the quadratic contrast for the sequence of utterance reached significance, F(1,16) =4.678, p = 0.046, with the vowel /a/ showing greater particulate emissions and the vowel /i/ showing the least. Given the relative mouth opening among /a/, /i/, and /u/ it appears as though mouth opening might be the reason for greater emissions. Descriptive statistics for particulate count for each speech sounds can be found in Table 2. Figure 6 shows a graphical comparison across categories.

**Figure 6. Number of particles per category**

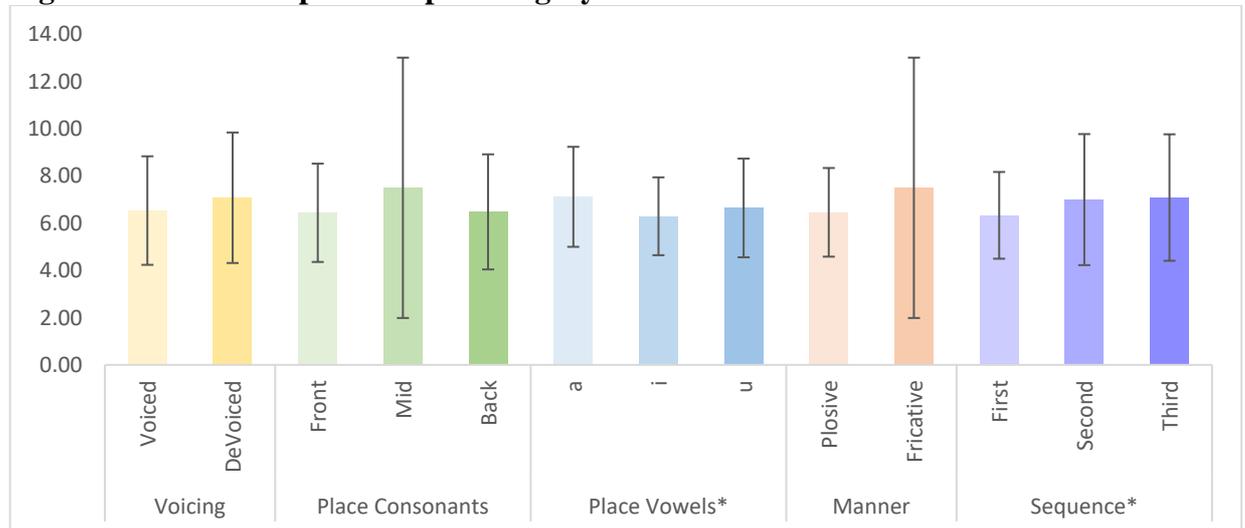

* Statistically significant comparison. Error bars represent standard deviation.

**Discussion and Conclusions**

This investigation sought to quantify aerosol emissions during the production of specific speech sounds that will serve as validation data for computational simulation and modelling of particulates during speech. Aerosols emitted via the oral cavity were captured using a time sequence of images with a laser and highspeed camera while participants produced different phonemes, representing specific linguistic qualities. The two purposes of this study were to test the feasibility of using a ND-YAG laser operating at 532 nm in a free field and to test if individual speech sounds differ in the number of particulates emitted from the oral cavity. This study served as a preliminary investigation to establish validation data for computations models of aerosol behavior during various speech tasks.

The results clearly demonstrate that the method of data collection can capture the increase in aerosolized particles due to human emission over the background dust, which was also quantified. One shortcoming included the lack of synchronization between the laser pulses and the camera frame rate during the recording of phoneme production. When the laser and camera were not precisely synchronized to optimize the laser illumination and shutter speed, the increase in particle counts observed during the vocal task may have been under-recorded. Further refinement to the methodology with higher resolution recordings and laser synchronization will be required to establish if there are significant differences in emission levels based on the phoneme produced. Despite this, the quantification of specific speech sounds above noise is apparent and will prove important in validating computational simulations and models. Additionally, the Micro-Particle Image Velocimetry (micro-PIV) technique can be utilized in future work to resolve the aerosols between 10 micron and 1 micron near the mouth aperture that were not likely resolved with the current PIV setup. Such quantification of the number, size, distribution, and speed of droplets being emitted within a centimetre of the mouth aperture (loosely defined as the near field) is critical in specifying an accurate boundary condition for numerical models of the polydisperse flow in the far field. Experimental quantification of the polydisperse flow in the near field using human participants may also enable parametric classification of currently less defined descriptors, such as super emitters and super spreaders.

In answering the second aim of this study, although there appear to be mixed results on whether specific speech sounds differ in the production of particulates, this method does appear to produce solid analysable data. Some phoneme classes showed clear differences, as in the case of the placement of vowels, while others did not. Although the method of speech sounds production could clearly show particulate emissions, establishing that there are the differences in particulate emission during specific phonemes production was not confirmed in this study. In addition to the potential for reduced particulate illumination due to the inconsistency in synchronization of images, it could be that our choice of phonemes did not adequately determine differences between phoneme classes. Additionally, it could be that comparisons across multiple phoneme categories, not simply one exemplar from each category is necessary to obtain significant and meaningful results. Determining the relative contribution of linguistic factors such as vocal fold vibration (voicing), the location in the vocal tract where airflow is manipulated and constricted (place), or how the airflow is manipulated (manner; completely or constricted occlusions) may be useful in establishing boundary conditions for computational models that would not only model aerosols emitted from the mouth but also model how those aerosols might be created in the vocal tract.

The lack of statistical power in this study might be another cause for failing to show differences between phonemes. Although this study analysed 17 healthy adults, the number of conditions and parameters that were studied might have weakened the power to observe the desired results. Although each participant produced each phoneme 3 times for a total of 54 observations per phoneme, this number might have fallen short. This is surprising given that other studies employing the same methods showed clear differences between their conditions. Mürbe and colleagues [19] employed only eight participants who produced five trials per condition. However, their analysis included 30 sec of data per trial analysing a larger time period. In addition, they compared running speech and singing, which may have yielded an overall greater difference between the tasks they were investigating. Good and colleagues [21] also compared singing vs. speaking tasks and employed 63 participants analysing data over a variable period of time. However, in this study they correlated particle count with other variables such as $CO_2$, which was not comparable for power analysis. In our study, our analysis only covered approximately 4 sec of data for three trials. Additionally, all conditions investigated in our study came from the same class of task, the production of isolated speech phonemes. Thus, power analysis for the number of trials per phoneme, or class of phonemes, should be investigated to unequivocally determine any differences across linguistic characteristics. Such discrimination will validate computational modelling with greater precision.

One statistically significant finding noted in this study revealed that the sequence of phoneme production differed between the first production and the third production. This finding is most likely due to the increase of particle accumulation across measures, which is likely why previous studies noted robust findings. The residence time of particles lingering was clearly longer than the time between phoneme productions. Measuring phoneme production with greater time between trials might result in no differences between initial and final phoneme production. This level of specificity might be important when validating a computational model but begs the question of ecological validity of the research design. It is highly unlikely that individuals produce single phonemes in daily communication, thus, the specificity needed for modelling validation might reduce the generalizability of research findings. Furthermore, phonemes are rarely produced in isolation and change according to the physical parameters of neighbouring phonemes. Coarticulatory factors will undoubtedly influence the nature and degree of particulate emissions. Continued research on not only single phonemes but also coarticulatory phenomena will reflect ecological situations and provide appropriate validation for models that predict how particles behave at the syllable, word, sentence, and conversational levels.

The wide range of variability in particle emissions among the participants was of particular note in this study. Within one phoneme category, two participants produced particle emissions 4 standard deviations from the mean and one other produced emissions at 2 standard deviations. The high degree of variability in particle count has been reported in two other studies,[16,17] which highlights the need for greater controls and more specific inclusion criteria in this type of study. The interparticipant variability also speaks to the inherent individual differences that might contribute to the aerosolization of mucous in the vocal tract, which is ultimately emitted from the oral cavity. Non-speech factors such as body mass index, age, total lung volume/capacity, systemic hydration level, medications, and mucosal tissue health are likely to contribute to the degree of particle emission. Additionally, controlling for speech intensity, speech kinematics and articulatory precision, respiratory drive, aerodynamic characteristics of vocal fold vibration, and vocal tract morphology will be important in developing a precise model of particle emission and behavior from the oral cavity.

This investigation characterized particulate emissions at the phonemic level in isolation, makes this study unique. The importance of understanding aerosolization at this level of specificity will be helpful in understanding *how* aerosols are emitted, not just *that* they are emitted. Knowledge of how particulates are emitted provides the necessary basis in the development of aerosol emission mitigation techniques. It also will prove useful in developing a more nuanced computational model that accounts for vocal tract configuration. A deeper understanding of how particulates are created, leave the oral cavity, and interact with their environment is necessary in order to develop targeted methods to mitigate the transfer of aerosolized viruses and other pathogenic material from one person to another.

**Acknowledgements**

The authors wish to thank Amy Nabors for her contribution in the development of the speech and singing stimuli, Hayleigh Wilson for her participation in participant runs, Brenda Hwang for scheduling participants, and Elizabeth McBride and Taylor Allay for their development of the counterbalancing procedures.

**Author Contributions**

Conceptualization:  Miriam van Mersbergen, Jeffrey Marchetta, Daniel Foti.

Data curation: Miriam van Mersbergen, Eric Alan Pillow, Chandler Cain, Stephen Morvant III.

Formal analysis: Miriam van Mersbergen, Jeffrey Marchetta, Daniel Foti, Eric Alan Pillow Apartim Dasgupta.

Writing – original draft: Jeffrey Marchetta, Miriam van Mersbergen, Daniel Foti

Writing – review & editing:

**References**

1. Jones RM, Brosseau LM. Aerosol transmission of infectious disease. *J Occup Environ Med*. 2015;57(5):501-508. doi:10.1097/JOM.0000000000000448
2. Bahl P, de Silva C, Bhattacharjee S, et al. Droplets and Aerosols Generated by Singing and the Risk of Coronavirus Disease 2019 for Choirs. *Clin Infect Dis*. 2021;72(10):e639-e641. doi:10.1093/cid/ciaa1241
3. Tellier R, Li Y, Cowling BJ, Tang JW. Recognition of aerosol transmission of infectious agents: A commentary. *BMC Infect Dis*. 2019;19(1):1-9. doi:10.1186/s12879-019-3707-y
4. Verreault D, Moineau S, Duchaine C. Methods for Sampling of Airborne Viruses. *Microbiol Mol Biol Rev*. 2008;72(3):413-444. doi:10.1128/mmbr.00002-08
5. Mittal R, Ni R, Seo JH. The flow physics of COVID-19. *J Fluid Mech*. 2020;894:1-14. doi:10.1017/jfm.2020.330
6. Sacha SB, Oliver BG, Blazey AJ, et al. Exhalation of Respiratory Viruses by Breathing, Coughing, and Talking. *J Med Virol*. 2009;81(9):1674-1679. doi:https://doi.org/10.1002/jmv.21556
7. Li Q, Guan X, Wu P, et al. Early Transmission Dynamics in Wuhan, China, of Novel Coronavirus–Infected Pneumonia. *N Engl J Med*. 2020;382(13):1199-1207. doi:10.1056/nejmoa2001316
8. Zou L, Ruan F, Huang M, et al. ARS-CoV-2 Viral Load in Upper Respiratory Specimens of Infected Patients. *N Engl J Med*. 2020;382:1177-1179.
9. Tan VYJ, Zhang EZY, Daniel D, et al. Respiratory droplet generation and dispersal during nasoendoscopy and upper respiratory swab testing. *Head Neck*. 2020;42(10):2779-2781. doi:10.1002/hed.26347
10. Naunheim MR, Bock J, Doucette PA, et al. Safer Singing During the SARS-CoV-2 Pandemic: What We Know and What We Don't. *J Voice*. 2021;35(5):765-771. doi:10.1016/j.jvoice.2020.06.028
11. Bourouiba L, Dehandschoewercker E, Bush JWM. Violent expiratory events: On coughing and sneezing. *J Fluid Mech*. 2014;745:537-563. doi:10.1017/jfm.2014.88
12. Tang JW, Settles GS. Coughing and Aerosols. *N Engl J Med*. 2008;359(15):e19. doi:10.1056/nejmicm072576
13. Tang JW, Liebner TJ, Craven BA, Settles GS. A schlieren optical study of the human cough with and without wearing masks for aerosol infection control. *J R Soc Interface*. 2009;6(SUPPL. 6):727-736. doi:10.1098/rsif.2009.0295.focus
14. Anderson EL, Turnham P, Griffin JR, Clarke CC. Consideration of the Aerosol Transmission for COVID-19 and Public Health. *Risk Anal*. 2020;40(5):902-907. doi:10.1111/risa.13500
15. Johnson TJ, Nishida RT, Sonpar AP, et al. Viral load of SARS-CoV-2 in droplets and bioaerosols directly captured during breathing, speaking and coughing. *Sci Rep*. 2022;12(1):1-13. doi:10.1038/s41598-022-07301-5
16. Asadi S, Wexler AS, Cappa CD, Barreda S, Bouvier NM, Ristenpart WD. Aerosol emission and superemission during human speech increase with voice loudness. *Sci Rep*. 2019;9(1):1-10. doi:10.1038/s41598-019-38808-z
17. Gregson FKA, Watson NA, Orton CM, et al. Comparing aerosol concentrations and particle size distributions generated by singing, speaking and breathing. *Aerosol Sci Technol*. 2021;55(6):681-691. doi:10.1080/02786826.2021.1883544
18. Coleman KK, Tay DJW, Tan K Sen, et al. Viral Load of Severe Acute Respiratory Syndrome Coronavirus 2 (SARS-CoV-2) in Respiratory Aerosols Emitted by Patients With Coronavirus Disease 2019 (COVID-19) While Breathing, Talking, and Singing.

*Clin Infect Dis*. 2021;2(Xx):1-7. doi:10.1093/cid/ciab691

19. Murbe D, Kriegel M, Lange J, Schumann L, Hartmann A, Fleischer M. Aerosol emission of adolescents voices during speaking, singing and shouting. *PLoS One*. 2021;16(2 February):1-10. doi:10.1371/journal.pone.0246819
20. Anfinrud P, Stadnytskyi V, Bax CE, Bax A, Stadnytskyi, V. Visualizing Speech-Generated Oral Fluid Droplets with Laser Light Scattering. *N Engl J Med*. 2020;382(21):2061-2063.
21. Good N, Fedak KM, Goble D, et al. Respiratory Aerosol Emissions from Vocalization: Age and Sex Differences Are Explained by Volume and Exhaled $CO_2$. *Environ Sci Technol Lett*. 2021;8(12):1071-1076. doi:10.1021/acs.estlett.1c00760
22. Alsved M, Matamis A, Bohlin R, et al. Exhaled respiratory particles during singing and talking. *Aerosol Sci Technol*. 2020;54(11):1245-1248. doi:10.1080/02786826.2020.1812502
23. Abkarian M, Mendez S, Xue N, Yang F, Stone HA. Speech can produce jet-like transport relevant to asymptomatic spreading of virus. *Proc Natl Acad Sci*. 2020;117:25237-25245.
24. Georgiou GP. Effect of different types of speech sounds on viral transmissibility: a scoping review. *Speech, Lang Hear*. December 2021:1-7. doi:10.1080/2050571X.2021.2014705
25. M. A, Stone HA. Stretching and break-up of saliva filaments during speech: A route for pathogen aerosolization and its potential mitigation. *Phys Rev Fluids*. 2020;5:102301.
26. Ahmed T, Wendling HE, Mofakham AA, et al. Variability in expiratory trajectory angles during consonant production by one human subject and from a physical mouth model: Application to respiratory droplet emission. *Indoor Air*. 2021:1896-1912.
27. Giovanni A, Radulesco T, Bouchet G, et al. Transmission of droplet-conveyed infectious agents such as SARS-CoV-2 by speech and vocal exercises during speech therapy: preliminary experiment concerning airflow velocity. . *Eur Arch Oto-Rhino-Laryngology*. 2021;279:1387-1692.
28. Gupta JK, Lin C-H, Chen Q. Characterizing exhaled airflow from breathing and talking. *Indoor Air*. 2010;20:31-39.
29. Hamada S, Tanabe N, Hirai T. Speech sounds and the production of droplets and aerosols. *Intern Med*. 2021;60:1649-1650.
30. Inouye S, Sugihara Y. Measurement of puff strength during speaking: comparison of Japanese with English and Chinese languages. *J Phonetic Soc Japan*. 2015;19:43-49.
31. Kusunose K, Matsunaga K, Yamada H, Sata M. Identifying the extent of oral fluid droplets on echocardiographic machine consoles in COVID-19 era. *J Echocardiogr*. 2020;18:268-270.
32. Tan ZP, Silwal L, Bhatt SP, Raghav V. Experimental characterization of speech aerosol dispersion dynamics. *Sci Rep*. 2021;11:3953. https://doi.org/10.1038/s41598-021-83298-7.
33. Titze IR. *Principles of Voice Production*. Prentice Hall; 1994. doi:science.1193125 [pii]\r10.1126/science.1193125
34. Mihaescu M, Khosla SM, Murugappan S, Gutmark EJ. Unsteady laryngeal airflow simulations of the intra-glottal vortical structures. *J Acoust Soc Am*. 2010;127(1):435-444. doi:10.1121/1.3271276
35. Švec JG, Titze IR, Popolo PS. Estimation of sound pressure levels of voiced speech from skin vibration of the neck. *J Acoust Soc Am*. 2005;117(3):1386-1394. doi:10.1121/1.1850074